\documentclass[%
 reprint,
 amsmath,amssymb,
 aps,
]{revtex4-1}

\usepackage{graphicx}
\usepackage{dcolumn}
\usepackage{xcolor}
\usepackage{soul}
\usepackage{here}
\usepackage{dcolumn}
\usepackage{bm}

\newcommand{\X}{{X}}
\newcommand{\Y}{{Y}}
\newcommand{\XY}{{XY}}

\newcommand{\pN}{\mathrm{pN}}
\newcommand{\nm}{\mathrm{nm}}

\newcommand{\mMol}{\mathrm{mM}}
\newcommand{\uMol}{\mathrm{\mu M}}

\newcommand{\pch}{\Pi_{ch}}
\newcommand{\pmc}{\Pi_{mc}}
\begin{document}


\title{Information flow, Gating, and Energetics in dimeric molecular motors}

\author{Ryota Takaki}
\affiliation{%
 Department of physics, The University of Texas at Austin, Austin, TX\\
}%


 \author{Mauro L. Mugnai and D. Thirumalai}
 \affiliation{Department of chemistry, The University of Texas at Austin, Austin, TX \\
 }%


\date{\today}

\begin{abstract}
Molecular motors belonging to the kinesin and myosin super family hydrolyze ATP by cycling through a sequence of chemical states. These cytoplasmic motors are dimers made up of two linked identical monomeric globular proteins. Fueled by the free energy generated by ATP hydrolysis, the motors walk on polar tracks (microtubule or filamentous actin) processively, which means that only one head detaches and executes a mechanical step while the other stays bound to the track.  Thus, the one motor head must regulate chemical state of the other, referred to as "gating", a concept that is not fully understood. Inspired by experiments,  showing that only a fraction of the energy from ATP hydrolysis is used to advance the kinesin motors against load, we demonstrate that additional energy is used for coordinating the chemical cycles of the two heads in the dimer - a feature that characterizes gating. To this end, we develop a general framework based on information theory and stochastic thermodynamics, and establish that gating could be quantified in terms of information flow between the motor heads. Applications of the theory to kinesin-1 and Myosin V show  that information flow occurs, with positive cooperativity,  at external resistive loads that are less than a critical value, $F_c$. When force exceeds $F_c$, effective information flow ceases. Interestingly, $F_c$, which is independent of the input energy generated through ATP hydrolysis, coincides with force at which the probability of backward steps starts to increase. Our findings suggest that transport efficiency is optimal only at forces less than $F_c$, which implies that these motors must operate at low loads under {\it in vivo} conditions. 
\end{abstract}

\pacs{Valid pACS appear here}
\maketitle


\section{\label{sec:level1}Introduction}
\begin{figure*}
\begin{center}
\includegraphics[width=0.8\textwidth]{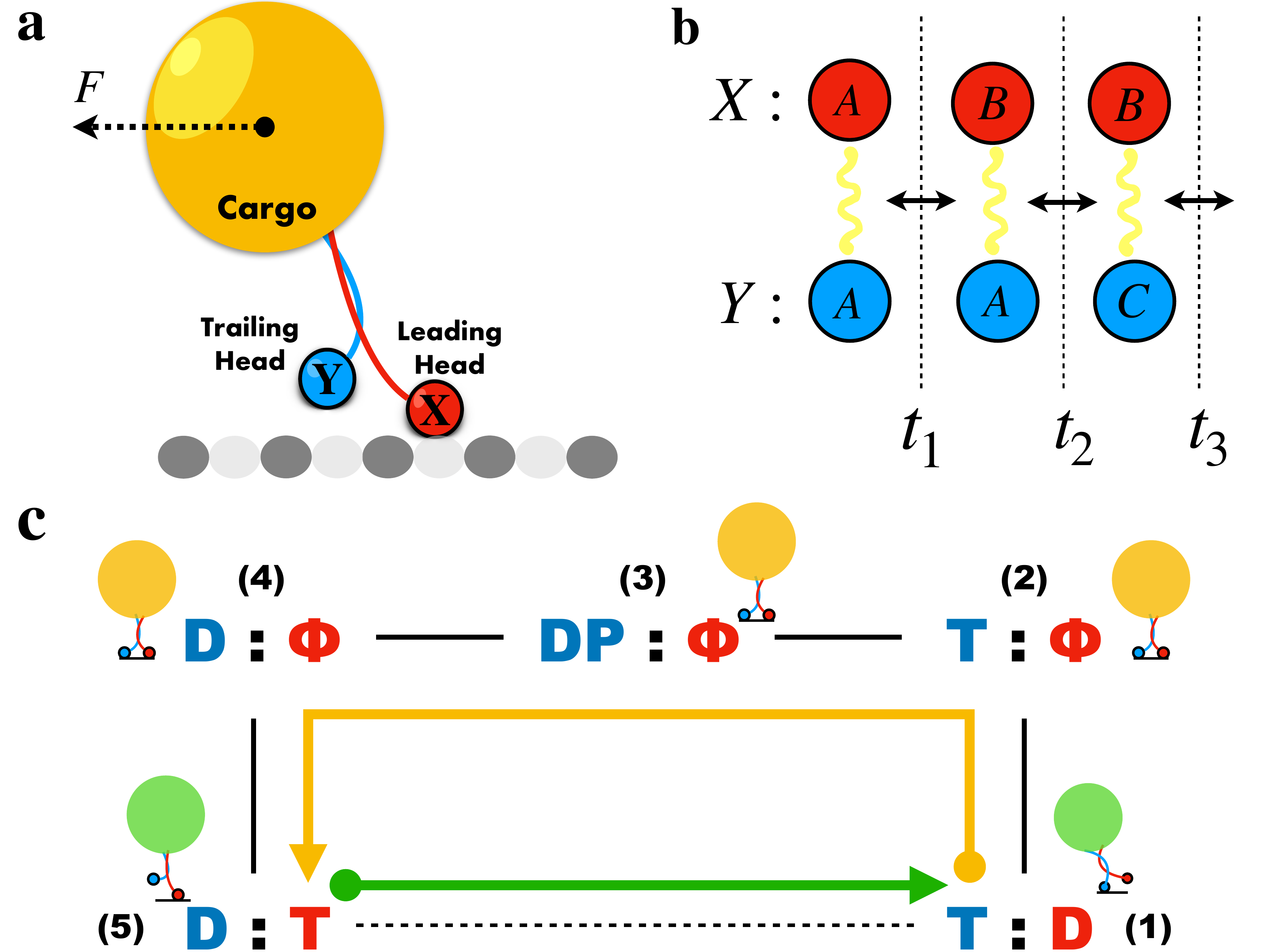}
\end{center}
\caption{\label{Fig:schematics} {\bf Kinesin stepping and associated transitions.} {\bf (a)} Schematics of  a typical optical trap experiment used to monitor stepping on the microtubule.  A resistive force, $F$,  acts on the cargo.   {\bf  (b)} Example of transitions for interacting systems with $ \X$ and $\Y$ being the leading  and trailing heads, respectively. The chemical states associated with $\X$ and $\Y$ are denoted by $A$,$B$, and $C$.  In this example, the transition in $\X$ is described as, $(x_0,y_0)=(A,A) \longrightarrow (x_1,y_0)=(B,A)$, and the transition in $\Y$ is given by $(x_1,y_0)=(B,A) \longrightarrow (x_1,y_1)=(B,C)$. Transitions occur at times $t_1$, $t_2$, $t_3,\cdots$.
{\bf (c)} Forward stepping cycle for kinesin. The  cycle is divided into chemical state (CS) transitions (yellow) and mechanical stepping (green). The chemical state of the trailing head is in blue, and that of the leading head is shown in red. The nucleotide state T denotes ATP, $\mathrm{\Phi}$ means  nucleotide state is not bound, D denotes ADP, and DP is the post hydrolyzed state with the trailing containing  ADP plus the inorganic phosphate, P$_i$. The caricatures for the physical states of the motor corresponding to each chemical state are displayed. }
\end{figure*}

Molecular motors utilize  chemical energy released by ATP hydrolysis in order to carry out multiple cellular functions that include transportation of vesicles~\cite{vale2000way,vale2003molecular,iino2020introduction}. Dimeric cytoplasmic motors (kinesin and myosin), constructed from two identical ATPases referred to as motor heads, walk on polar tracks (F-actin or microtubule, MT)  by a hand-over-hand mechanism (Fig.\ref{Fig:schematics}a)~\cite{Asbury2003Science,Yildiz2003Science,Yildiz2004Science}. In order for dimeric motors to take multiple steps in the forward direction, without disengaging from the polar track, there has to be coordination or communication between the motor heads.  This implies that the trailing head (TH) should detach with substantially higher probability than the leading head (LH) (a manifestation of inter-head communication)  while the LH should remain strongly bound to the polar track.  Because the nucleotide state (for example ATP-bound or ADP-bound) of the motor heads dictates the affinity for the polar track it follows that the ATPase cycle in the TH and LH should be partially out of phase~\cite{ma1997interacting} to ensure effective inter-head communication. In kinesin-1, ATP binds to the LH only after the ADP bound TH detaches from the MT~\cite{Dogan2015CellRep,Isojima2016NatChemBio}.  In myosin V, inter-head communication results from faster ADP release from the TH than from the LH.~\cite{Sakamoto2008Nature,Kodera2010Nature} Thus, chemical state regulation of one motor head by the other plays a significant role in the ability of the motor to take multiple steps on the polar track. This is referred to as ``gating", which may be viewed as a form of allosteric regulation~\cite{thirumalai2019symmetry}, in molecular motors. Gating, which is necessary for dimeric motors to maintain processivity, is possibly mediated by inter-head mechanical strain through the structural elements connecting the two heads~\cite{Hancock_1999,YILDIZ20081030,Hyeon_2007,SHASTRY2010939,Zhang12Structure,clancy2011universal,andreasson2015examining,milic2014kinesin,rosenfeld2003stepping,toprak2009kinesin,miyazono2010strain,hinczewski2013design}. 
Decrease in the gating efficiency in artificial constructs that mutate these elements (for example elongation of  neck linker in conventional kinesin)  results in  the display reduction of velocity and run length, and a decrease in the stall force~\cite{andreasson2015examining,YILDIZ20081030,miyazono2010strain,clancy2011universal}.
From these observations, we surmise that communication between the dimeric motors must be directly linked to energy costs needed to drive stepping. 


We develop a framework, based on stochastic network models, which are used to  develop a theory to quantify gating in terms of information theory.
The relevance of information flow between subsystems in the context of biology has been recognized in  recent developments that connect information and thermodynamics~\cite{parrondo2015thermodynamics,Sagawa_2009,horowitz2014thermodynamics}. For example, the significance of information transduction in the adaptive sensory system has been studied~\cite{ito2015maxwell,sartori2014thermodynamic}. Chemical nanomachines were investigated from information-thermodynamic perspective~\cite{amano2021insights,loutchko2017stochastic}.

We were motivated to cast gating in light of  information theory because astounding developments in experimental techniques have made it possible to measure  energetics in molecular motors with high accuracy~\cite{Toyabe_2010,PhysRevLett.121.218101}.  In a recent single molecule experiment, Ariga {\it et al.}~\cite{PhysRevLett.121.218101} measured energetics of kinesin by attaching a probe to the motor. The experiments produced two important results. (1)  They established that the motor functions out of equilibrium by showing that the relation between response and correlation function, expected for a system at equilibrium, does not hold.   The  extent of violation of the fluctuation response relation was quantified by using the Harada-Sasa equality~\cite{harada2005equality,harada2006energy}. (2) Based on the experimental data and a theoretical model, it was suggested~\cite{PhysRevLett.121.218101}  that the total heat dissipated is $\approx$80\% of the input energy, which implies that only about $\approx$20\% of the energy  from ATP hydrolysis is utilized in performing work. We surmised that part of the energy generated by ATP hydrolysis is invested in the coordination of chemical states between the two motor heads. Because coordination is the essence of gating, this can be interpreted and quantified as the energetic cost of gating itself.  

Our  theory  is based on stochastic network models. We assume that in the process stepping along the cytoskeletal filament, the dimeric motor could be driven into a non-equilibrium steady state (NESS) \cite{hwang2018energetic,hwang2016quantifying,ge2010physical,seifert2012stochastic,liepelt2007steady,mugnai2020theoretical,mugnai2020processivity}, provided that it does not detach from the polar track. The present work, which provides a framework for understanding gating in cytoplasmic motors in terms of information flow, utilizes and builds  on the most insightful formulation by Horowitz and Esposito~\cite{horowitz2014thermodynamics}.  In the process, we establish precise connections between inter-head communication and  energetics in dimeric motors.  Although we only consider stepping in kinesin-1 (Kin-1) and Myosin V, our approach is general, and could be used to understand gating and energy costs in other molecular machines~\cite{mugnai2020theoretical}. We show that the energy costs involved in the operation of the motors is related to both chemical state transitions and  information flow (gating) between the motor heads.


\section{Theory}
\subsection{\label{}Notation and assumption}
We consider two subsystems, labelled $\X$ and $\Y$~\cite{horowitz2014thermodynamics}, which in the context of molecular motors represent the leading head (LH) and the trailing head (TH), respectively. Each subsystem is specified by microstates $x$ and $y$ (see Fig.\ref{Fig:schematics}b). A pair of states associated with $\X$ and $\Y$ is $(x,y)$.   
We restrict ourselves to a network connecting $(x,y)$ that are bipartite, which means that only $\X$ or $\Y$ can alter  its state in a single transition. In other words, the transitions $(x_1,y_1)\rightarrow(x_2,y_1)$ and $(x_1,y_1)\rightarrow(x_1,y_2)$ are allowed but not $(x_1,y_1)\rightarrow(x_2,y_2)$ with $x_1 \neq x_2$ and $y_1 \neq y_2$. However, we require that the bipartite structure only along the path in which we calculate the interaction between $\X$ and $\Y$. 
Non-bipartite feature, which is needed in describing stepping in molecular motors,  is allowed otherwise. 

The transition rate for $(x_i,y_i) \rightarrow (x_{i+1},y_i)$ (and $(x_i,y_i) \rightarrow (x_{i},y_{i+1})$) is denoted as  \small$w_{x_ix_{i+1}}^{y_i}$\normalsize (\small$w_{x_i}^{y_iy_{i+1}}$\normalsize). The steady state joint probability for the state $(x,y)$ is  $p(x,y)$,  and the marginal probabilities for $\X$ and $\Y$ are $p_\X(x)=\sum_{y}^{}p(x,y)$ and $p_\Y(y)=\sum_{x}^{}p(x,y)$, respectively.
We describe the steady state properties of the motor. Thus, the probabilities are appropriate steady state quantities  in the stochastic kinetic networks for the motors. We set the temperature $T$ to unity unless specified.

\subsection{\label{}Information along transitions}
The point-wise mutual information (PMI), which plays an important role in this study, in the energy unit, $i(x,y)$, for a state $(x,y)$ is, 
\begin{align} 
\begin{split}
\label{eq:i_def}
i(x,y)=k_B\ln \frac{p(x,y)}{p_{\X}(x)p_{\Y}(y)},
\end{split} 
\end{align}
where $k_B$ is the Boltzmann constant. It is necessary to introduce PMI, instead of considering only the Mutual Information (MI), because information flow in motors requires quantifying the microscopic transitions in each step in the catalytic cycle (Fig.\ref{Fig:schematics}c). 
The MI is the average over all the states, $\sum_{x,y} p(x,y)i(x,y)$.
 
Let $\X$($\Y$) undergo $n_\X$($n_\Y$) transitions along a specific path. The difference in the PMI between the initial and final state associated with the path, $\Delta i$, can be calculated as,  $\Delta i_X +  \Delta i_Y$, where, 
\begin{align} 
\begin{split}
\label{}
\Delta i_\X&=\sum_{i=0}^{n_\X-1}[i(x_{i+1},y_{i})-i(x_{i},y_{i})] \\ 
&= k_B \ln\prod_{i=0}^{n_{\X}-1}\frac{p(y_{i}|x_{i+1})}{p(y_i|x_i)},
\end{split} 
\end{align}
where $p(y_i|x_i) = p(x_i,y_i)/p_X(x_i)$ is a conditional probability.
A similar expression holds for $\Delta i_\Y$. 
The bipartite property allows us to count the transitions for $\X$ and $\Y$ separately. Thus, $(x_i,y_i)$ means the state of $\X$ and $\Y$ before the $(i+1)^{th}$ transition for $\X$, and similarly $(x_j,y_j)$ for $\Y$ (see Fig.\ref{Fig:schematics}b).
The change in PMI along a path (for example (1)$\rightarrow$(2) transition in Fig.\ref{Fig:schematics}c), $\Delta i$, is given by,
\begin{align} 
\begin{split}
\label{eq:MI}
\Delta i&=k_B\ln\prod_{i=0}^{n_{\X}-1}\frac{p(y_{i}|x_{i+1})}{p(y_i|x_i)}\prod_{j=0}^{n_{\Y}-1}\frac{p(x_{j}|y_{j+1})}{p(x_j|y_j)}.
\end{split} 
\end{align}
where $\Delta i$ is the net information transferred between $\X$ and $\Y$. We are interested in how the input energy, arising from ATP hydrolysis, is parsed into the ``invisible" Chemical State (CS)  transitions [from (1) to (5) in the catalytic cycle in Fig.\ref{Fig:schematics}c], and mechanical transition [(5)$\rightarrow$(1) in Fig.\ref{Fig:schematics}c]. Therefore,  it is the information flow along a specific path (for instance, [(2)$\rightarrow$(3) in Fig.\ref{Fig:schematics}c], rather than information flow in the closed cycle (transition starting in (1) and ending in (1), resulting in the LH and TH switching positions on the polar track)  that is relevant for gating.
In contrast, for different reasons previous studies~\cite{horowitz2014thermodynamics,hartich2014stochastic,cafaro2016thermodynamic}, focused on the information flow  using the MI. Therefore, information flow of interest in these studies are restricted to closed cycle in which all the edges in the network are bipartite. As a result the net information transfer between two subsystems is zero in NESS. In Eq.(\ref{eq:MI}), the net information flow between two subsystems need not be zero along a specific path but vanishes  only when we compute the information flow for a closed path.
In molecular motors, we need to quantify heat dissipation along the invisible CS transitions in order to decipher not only the efficiency of the motor but also the energy cost associated with gating. For this reason, we calculated path or transition specific energy cost for a fixed input energy.

\subsection{\label{}Entropy production along the transitions}
Entropy production, $\rho$, along a path in steady state can be decomposed as~\cite{Seifert_2012}, 
\begin{align} 
\begin{split}
\rho = Q + \Delta s_\XY,
\end{split} 
\end{align}
where
\begin{align} 
\begin{split}
\label{Eq:EntProd}
Q&= k_B \ln \prod_{i=0}^{n-1}\frac{w_{x_i x_{i+1}}^{y_{i}y_{i+1}}}{w_{x_{i+1} x_{i}}^{y_{i+1}y_{i}}}
\end{split} 
\end{align}
is the dissipated heat and,
\begin{align} 
\begin{split}
\Delta s_\XY &=k_B \ln \prod_{i=0}^{n-1} \frac{p(x_i,y_i)}{p(x_{i+1},y_{i+1})}
\end{split} 
\end{align}
is the change of the entropy for the system.  
In Eq.(\ref{Eq:EntProd}), $w_{x_i x_{i+1}}^{y_{i}y_{i+1}}$ \normalsize is the transition rate from state $(x_i,y_i)$ to $(x_{i+1},y_{i+1})$. Entropy production is always positive along the direction of stationary flow because \small $p(x_i,y_i)w_{x_i x_{i+1}}^{y_{i}y_{i+1}}-p(x_{i+1},y_{i+1})w_{x_{i+1} x_{i}}^{y_{i+1}y_{i}}>0$ \normalsize for all $i$. For a bipartite graph,  $\rho$ can be decomposed as, 
\small
\begin{align} 
\begin{split}
\rho = \rho_\X + \rho_\Y = (Q_\X + \Delta s_\XY^\X) + (Q_\Y + \Delta s_\XY^\Y),
\end{split} 
\end{align}
\normalsize
where 
\begin{align} 
\begin{split}
Q_\X &= k_B \ln \prod_{i=0}^{n_\X-1}\frac{w_{x_i x_{i+1}}^{y_{i}}}{w_{x_{i+1} x_{i}}^{y_{i}}}; \Delta s_\XY^\X = k_B\ln \prod_{i=0}^{n_\X-1} \frac{p(x_i,y_i)}{p(x_{i+1},y_{i})};  \\ 
 Q_\Y &= k_B \ln \prod_{j=0}^{n_\Y-1}\frac{w_{x_j}^{y_{j}y_{j+1}}}{w_{x_{j}}^{y_{j+1} y_{j}}}; \Delta s_\XY^\Y = k_B\ln \prod_{j=0}^{n_\Y-1} \frac{p(x_j,y_j)}{p(x_{j},y_{j+1})}.  \\ 
\end{split} 
\end{align}
We can show that the equalities, $\Delta s_\X = \Delta s_\XY^\X + \Delta i_\X$ and $\Delta s_\Y = \Delta s_\XY^\Y + \Delta i_\Y$ hold,
where $\Delta s_\X=k_B\ln \prod_{i=0}^{n_\X-1} \frac{p_\X(x_i)}{p_\X(x_{i+1})}$ and $\Delta s_\Y=k_B\prod_{j=0}^{n_\Y-1}\ln \frac{p_\Y(y_i)}{p_\Y(y_{i+1})}$.
Using these relations, we obtain for each $\X$ and $\Y$, 
\begin{align} 
\begin{split}
\label{Eq:equality_rho_s_i}
\rho_{\X(\Y)}
&=\sigma_{\X(\Y)} -\Delta i_{\X(\Y)}. \\ 
\end{split} 
\end{align}
In Eq.~(\ref{Eq:equality_rho_s_i}), we defined the apparent entropy production from $\X$ and $\Y$ as $\sigma_\X \equiv  Q_\X + \Delta s_\X$ and $\sigma_\Y \equiv  Q_\Y + \Delta s_\Y$, respectively. By summing the terms in  Eq.(\ref{Eq:equality_rho_s_i}) for $\X$ and $\Y$, we obtain,
\begin{align} 
\begin{split}
\label{Eq:Epprod}
\rho = \sigma - \Delta i,
\end{split} 
\end{align}
where $\rho \equiv \rho_\X + \rho_\Y$ and $\sigma \equiv \sigma_\X + \sigma_\Y$. 
The values of $\rho$, $\rho_\X$, and $\rho_\Y$, representing entropy production,  are always positive in the direction of stationary flow. Therefore, the lower bounds for the apparent entropy production, $\sigma$, $\sigma_\X$, and $\sigma_\Y$ are $\Delta i$, $\Delta i_\X$, and $\Delta i_\Y$, respectively. The value of $\sigma$ ($\sigma_\X$ and $\sigma_\Y$ for the subsystems) is allowed to be negative provided $\Delta i \le \sigma$.  
Negative values might be indicative of an apparent violation of the second law of thermodynamics, and is sometimes taken to be a signature of Maxwell demon~\cite{ito2013information,esposito2012stochastic,sagawa2013role,horowitz2014thermodynamics}. We discuss the Maxwell demon regime (MDR) further below.

\subsection{\label{} Free energy transduction and information}
We consider an isothermal environment at temperature  $T$ with the possibility that  material exchange occurs resulting in the break down of the detail balance condition. In the context of molecular motors, this is realized by having ligands [ATP, ADP, phosphate (P$_i$)] that are maintained at constant concentrations~\cite{ge2010physical}.  In the NESS,  it is possible to identify the entropy production as free energy transduction ($\Delta \mu$) along the path~\cite{zhang2012stochastic}, $\Delta \mu = \rho$.  It is obvious from the previous section that the transduction is written using the apparent free energy dissipation  and the associated information flow terms:
\begin{align} 
\begin{split}
\label{Eq:mainEq}
\Delta \mu = \sigma - \Delta i.
\end{split} 
\end{align}
The equation above relates the chemical energy supplied from ATP hydrolysis (ATP$\rightarrow$ADP + P$_i$) and information operation associated with coordination between the two heads in the dimeric motors. Although hydrolysis occurs in a single step [(2)$\rightarrow$(3) in Fig.\ref{Fig:schematics}c], it is a multi-step enzyme reaction (ATP binding, hydrolysis, and release of P$_i$ and ADP) that impacts the chemical steps in the catalytic cycle. It is worth mentioning that some of the energy is consumed in driving the chemical transitions in the catalytic cycle of the motor [(1)$\rightarrow$(5) in Fig.\ref{Fig:schematics}c]. The decomposition given in Eq.~(\ref{Eq:mainEq}) is particularly useful in molecular machines in which the mechanical function is not coupled to chemical reactions. This is indeed the case in  kinesin as well as Myosin V, as we show below.

\subsection{\label{}Mechanochemistry and Information}
 We first illustrate the connection between $\Delta i$ and motor movement using kinesin as an example.
We model the forward stepping cycle for kinesin using the kinetic network shown in Fig.\ref{Fig:schematics}c. Completion of a single step (cycle) begins by releasing ADP from the LH  [$(1)\rightarrow(2)$], followed by ATP hydrolysis in the TH  [$(2)\rightarrow(3)$],  release of P$_i$ [$(3)\rightarrow(4)$],  and ATP binding to the leading head [$(4)\rightarrow(5)$]. The four chemical transitions poises the TH to take a mechanical step [$(5)\rightarrow(1)$]. The catalytic cycle in Fig.\ref{Fig:schematics}c naturally decomposes into CS transitions [state (1) to state (5)] ($\pch$) that alter the nucleotide states of the motor, and the mechanical stepping [state (5) to state (1)] ($\pmc$). Once mechanical stepping is complete,  the next cycle of chemical transitions is initiated. 
In the following analysis, we write the probability in state $(i)$ as $p_i$ and the rate of transition from state $(i)$ to state $(j)$ as $w_{ij}$.

In anticipation of the link between information flow and gating, we begin with energy conservation associated with the cycle in Fig.\ref{Fig:schematics}c.
The motor operates by input of energy, $\mu$, arising from ATP hydrolysis.
We partition this energy into two contributions: $\mu_{ch}$, associated with the chemical transitions $\Pi_{ch}$, and the energy $\mu_{mc}$ expended in the mechanical step $\Pi_{mc}$, which advances the motor along the microtubule by one step ($\approx 8$ nm for kinesin).
Conservation of energy implies that,
\begin{equation}
\mu = \mu_{ch} + \mu_{mc}.
\label{Eq:EnergyCons}
\end{equation}
In terms of microscopic transition rates, we can write,
\begin{equation}
\label{Eq:affi}
\mu = k_B \ln \frac{w_{12}w_{23}w_{34}w_{45}w_{51}}{w_{21}w_{32}w_{43}w_{54}w_{15}} + W,
\end{equation}
where $W$ is the mechanical work done by the motor. We define,
\begin{equation}
\mu_{ch} = k_B \ln \frac{w_{12}w_{23}w_{34}w_{45}}{w_{21}w_{32}w_{43}w_{54}} + k_B \ln \frac{p_1}{p_5}.
\label{Eq:much}
\end{equation}
and,
\begin{equation}
\mu_{mc} = k_B \ln \frac{w_{51}}{w_{15}} + k_B \ln \frac{p_5}{p_1} + W.
\label{Eq:mume}
\end{equation}

Using Eq.(\ref{Eq:mainEq}) for $\pch$ we obtain, 
\begin{align} 
\begin{split}
\label{Eq:mu_chem}
\mu_{ch}=\sigma_{ch}-\Delta i_{ch},\\ 
\end{split} 
\end{align}
where $\sigma_{ch}$ and $\Delta i_{ch}$ is the apparent entropy production and the change of PMI in $\pch$, respectively.
For $\pmc$ the following equation holds:
\begin{align} 
\begin{split}
\label{Eq:mu_mec}
\mu_{mc}-W=\rho_{mc},\\ 
\end{split} 
\end{align}
where $W$ is the mechanical work done by the motor and $\rho_{mc}=k_B\log\frac{w_{51}p_5}{w_{15}p_1}$ is the entropy production in $\pmc$. $\rho_{mc}$ quantifies the non-equilibrium nature of the motor stepping, in other words, $\rho_{mc}=0$ at equilibrium, which is a consequence of the detailed balance condition in Fig.~\ref{Fig:schematics}c). The sum of Eq.(\ref{Eq:mu_chem}) and Eq.(\ref{Eq:mu_mec}) results in $\mu-W=Q$ which is the energy conservation for the whole cycle. This is explicitly demonstrated in Fig.\ref{Fig:CErate} a-b for kinesin.

Because $\mu_{mc}$ is given by $\mu-\mu_{ch}$, 
it follows that
\begin{align} 
\begin{split}
\label{Eq:Infgain}
\mu_{mc}=\mu + \Delta i_{ch} -\sigma_{ch}.\\ 
\end{split} 
\end{align}
It can be shown that \small $\text{e}^{(\mu_{mc}-W)/k_B}= (p_5 w_{51})/ (p_1 w_{15})$ \normalsize holds for the cycle in Fig.\ref{Fig:schematics}c. Using the flux associated with the mechanical step, $J_{mc}=p_5 w_{51}-p_1 w_{15}$, we obtain,
\begin{align} 
\begin{split}
\label{Eq:flux}
J_{mc}=p_1 w_{15}(\text{e}^{(\mu_{mc}-W)/k_B}-1) .
\end{split} 
\end{align}
Expressions for the key quantities,  such as $\Delta i_{ch}$, for the kinetic network in Fig.\ref{Fig:schematics}c are given in the Appendix~\ref{sec:expressions}. 

From Eq.(\ref{Eq:Infgain}) and Eq.(\ref{Eq:flux}), it follows that when $\Delta i_{ch}>0$, the flux, $J_{mc}$ in $\pmc$ increases. This implies that by increasing $\Delta i_{ch}$, more energy ($\mu_{mc}$) would be available during the mechanical step [see Eq.(\ref{Eq:Infgain})].  
Therefore, positive $\Delta i_{ch}>0$ implies that the communication between the heads is efficient, a situation that we refer to as positive cooperativity.  In contrast, when $\Delta i_{ch}<0$ (negative cooperativity) the motor has to expend  $\Delta i_{ch}$ in addition to $\sigma_{ch}$ to execute the chemical transitions in order to complete the cycle [see Eq.(\ref{Eq:mu_chem}) and (\ref{Eq:Infgain})]. This results in a decrease in the available energy for carrying out the mechanical step.  Therefore,  we surmise that information flow {\it $\Delta i_{ch}$ quantifies the efficacy of gating in dimeric motors.} 


\section{Kinesin}
\begin{figure}
\centering
\includegraphics[width=0.48\textwidth]{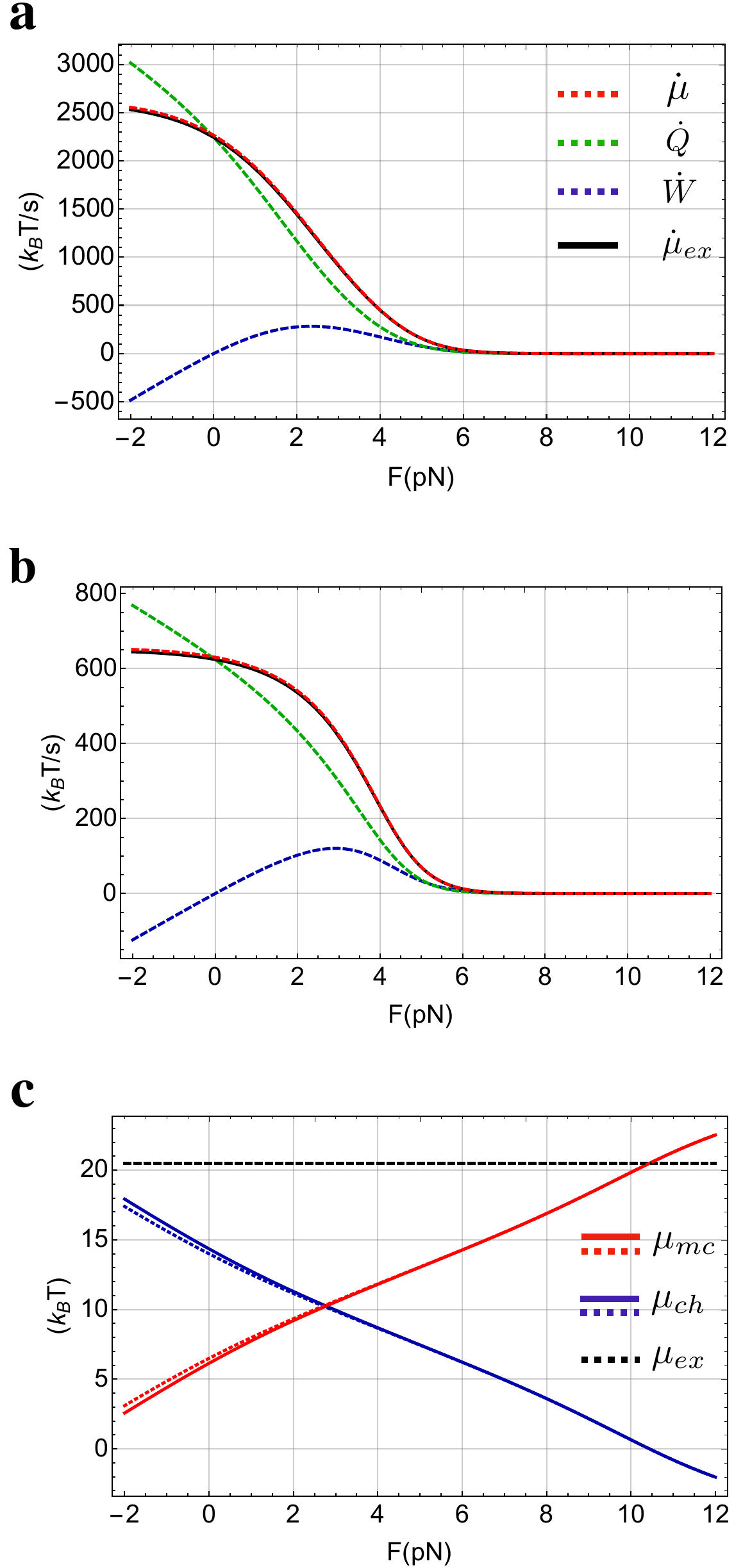}
\caption{\label{Fig:CErate} {\bf(a)-(b)}  Rates of energy expenditure  in kinesin as a function of the external load, $F$. The rates of work (power) performed by the motor and the rate of heat dissipation calculated from the kinetic network in Fig.~\ref{Fig:schematics}c are denoted as $\dot{W}$ and $\dot{Q}$, respectively. The sum of $\dot{W}$ (blue dashed line) and $\dot{Q}$ is $\dot{\mu}$ (green). The rate of input energy in the single molecule experiment~\cite{PhysRevLett.121.218101} (black) is $\dot{\mu}_{ex}$. {\bf (a)} [T]=1mM, [D]=0.1mM, and [P$_i$]=1mM. {\bf (b)} [T]=10$\mu$M, [D]=1$\mu$M, and [P$_i$]=1mM. Under both the conditions $\dot{\mu} \equiv \dot{\mu}_{ex}$, which not only validates the theory but also shows that the extracted parameters for kinesin (see TABLE I) using different observables are reasonable. {\bf (c)} Allocation of the input chemical energy for mechanical transition ($\mu_{mc}$) and chemical transitions ($\mu_{ch}$) as a function of $F$. The results are obtained theoretically for the kinesin network in Fig.~\ref{Fig:schematics}c . The black dashed line shows the experimental value, $\mu_{ex}$, which is the energy input.  Solid lines are for \{[T]=1mM, [D]=0.1mM, [P$_i$]=1mM\} and dotted lines are for \{[T]=10$\mu$M, [D]=1$\mu$M, [P$_i$]=1mM\}. The calculated sum, $\mu_{ch} + \mu_{mc}$, is equal to $\mu_{ex}$. }
\end{figure}

By using our theory, we first analyzed the energetics and information flow in kinesin-1 using the experimental data~\cite{PhysRevLett.121.218101}. The implementation of the theory requires the rates of various transitions in the catalytic cycle of the motor. We follow previous studies~\cite{liepelt2007kinesin,hwang2018energetic} to extract the kinetic rates in our model (Fig.\ref{Fig:schematics}c). Detailed description of the procedure to calculate the parameters and their values  are in the Appendix~\ref{sec:parameters}. 

Because our theory is based on thermodynamics, it is important to assess if the model in Fig.\ref{Fig:schematics}c reflects the energetics in kinesin in the experiment~\cite{PhysRevLett.121.218101}. Specifically, the input energy needed to drive the catalytic cycle in the experiment ($\mu_{ex}$) has to equal to the sum of dissipated heat ($Q$) and work done by kinesin ($W$) calculated using the network shown in Fig.\ref{Fig:schematics}c. In Figs.\ref{Fig:CErate}a-b, we plot the rate for $\mu_{ex}$ and $\mu=Q+W$, obtained by multiplying each quantity by the steady state current in the network using the parameters listed in Table I.  The agreement between the calculations and experiment is excellent for both two sets of nucleotide concentrations. Thus, we conclude that the network used in the analysis describes the thermodynamics of energy consumption in kinesin accurately.

It is worth noting that the power generated, $\dot{W} = FV$ ($V$ is the motor velocity) reaches a maximum value at $F \ne 0$ (see Figs. \ref{Fig:CErate}a-b) because $V$ is a decreasing function of $F$, and hence the product has a maximum. The location of the maximum depends on the concentrations, [T], [D], and [P$_i$] (compare Figs. \ref{Fig:CErate}a and b). At the maximum of $\dot{W}$, the ratio $\dot{W}/\dot{\mu} \approx 0.2$, which nearly coincides with the estimate in the experiment~\cite{PhysRevLett.121.218101} using a combination of measurements and a mathematical model. 

Two different nucleotide concentrations, \{[T]=1mM, [D]=0.1mM, [P]=1mM\} and \{[T]=10$\mu$M, [D]=1$\mu$M, [P]=1mM\} were used in the experiments~\cite{PhysRevLett.121.218101}. Under both the conditions  the input  energy ($\mu_{ex}=84.5\ \pN \cdot \nm\approx20.5\ k_B T $)   is the same because  [T]/([D][P]) is identical. 
The ideal stall load ($F_{max}$) may be estimated using,  $F_{max}=20.5(k_BT)\cdot4.1(\nm \cdot \pN/k_BT)/8(\nm)=10.5\pN$, where 8nm is the step size of kinesin.  However, the measured stall load ($F_s$) for kinesin-1 is in the range $6\pN \lesssim F_s \lesssim 8\pN$~\cite{svoboda1994force,visscher1999single,block2003probing}. The reason for this is that pathways, besides the forward steps, become relevant when $F$ exceeds a critical value (discussed further below).

The mechanisms of stepping kinetics in motors are usually inferred by  applying an external load to the cargo (Fig.\ref{Fig:schematics}a) in single molecule optical tweezer experiments. In order to  the extract the motor energetics from such experiments, one has to be cautious because heat dissipation occurs in a multidimensional energy landscape, while experiments only report the dynamics  along one experimentally accessible dimension, stepping direction, which is aligned with $F$.  This is clear because a source of hidden dissipation, which can not be accessed by observing only the steps, is the energy ($Q_{ch}$) expended to drive the chemical transitions. Thus, only a portion of the input energy ($\mu$) is used for mechanical stepping.  We calculated $\mu_{ch}$ and  $\mu_{mc}$ as a function of $F$ (Fig.\ref{Fig:CErate}c) in order to determine the allocation of the input energy ($\mu$), at the two different nucleotide concentrations used in the experiment, for mechanical stepping ($\mu_{mc}$), and chemical transitions ($\mu_{ch}$). The values of $\mu_{mc}$ and $\mu_{ch}$ are almost identical under both the conditions, suggesting that the allocation does not depend on the nucleotide concentrations.
From the plots in Fig.\ref{Fig:CErate}c, we find  that $\mu_{mc}\approx9.2\ k_BT$ and $\mu_{ch}\approx11.3\ k_BT$ at $F=2\pN$, the value of resistive force used to estimate heat dissipation kinesin in the experiment~\cite{PhysRevLett.121.218101}. Thus, at most 45 \% of the input energy is used to drive mechanical stepping at $F=2\pN$. As $F$ increases, the energy needed to drive the mechanical step increases, which must come at the expense of a decrease in the $\mu_{ch}$ for executing the chemical transitions. 
Although theoretically, the entire input energy could be used for stepping at $F=10.5\pN $ (Fig.\ref{Fig:CErate}c) only by maintaining the forward cycle at equilibrium, it cannot be realized at all because CS transitions that cost energy have to occur for the motor to step forward. 
It also follows that the  ideal stall force $F_{max}=10.5\pN$, which cannot be obtained in experiments because at moderately high forces ($F>5\pN$) backward steps start to be prominent. 

The discrepancy between $F_{max}$ and $F_s$ is a consequence of the non-equilibrium nature of the stepping transition, which creates alternate stepping pathways that consume energy. The kinetic diagram, at loads $F>5\pN$, is best described using a dual cycle  because there is an increase in the probability of backward steps~\cite{liepelt2007kinesin,hwang2018energetic}. The stall of kinesin is a consequence of the dynamic equilibrium between the cycles for forward and  backward stepping.  It is important to note that backward steps start to be prominent, typically at $\approx 5 \pN$~\cite{visscher1999single,hwang2018energetic} in kinesin, leading to the experimentally observed value of $F_s$ that is considerably less than $F_{max}$. We explain this finding by calculating $\Delta i_{ch}$, quantification of the cooperativity.

\begin{figure*}[]
\centering
\includegraphics[width=\textwidth]{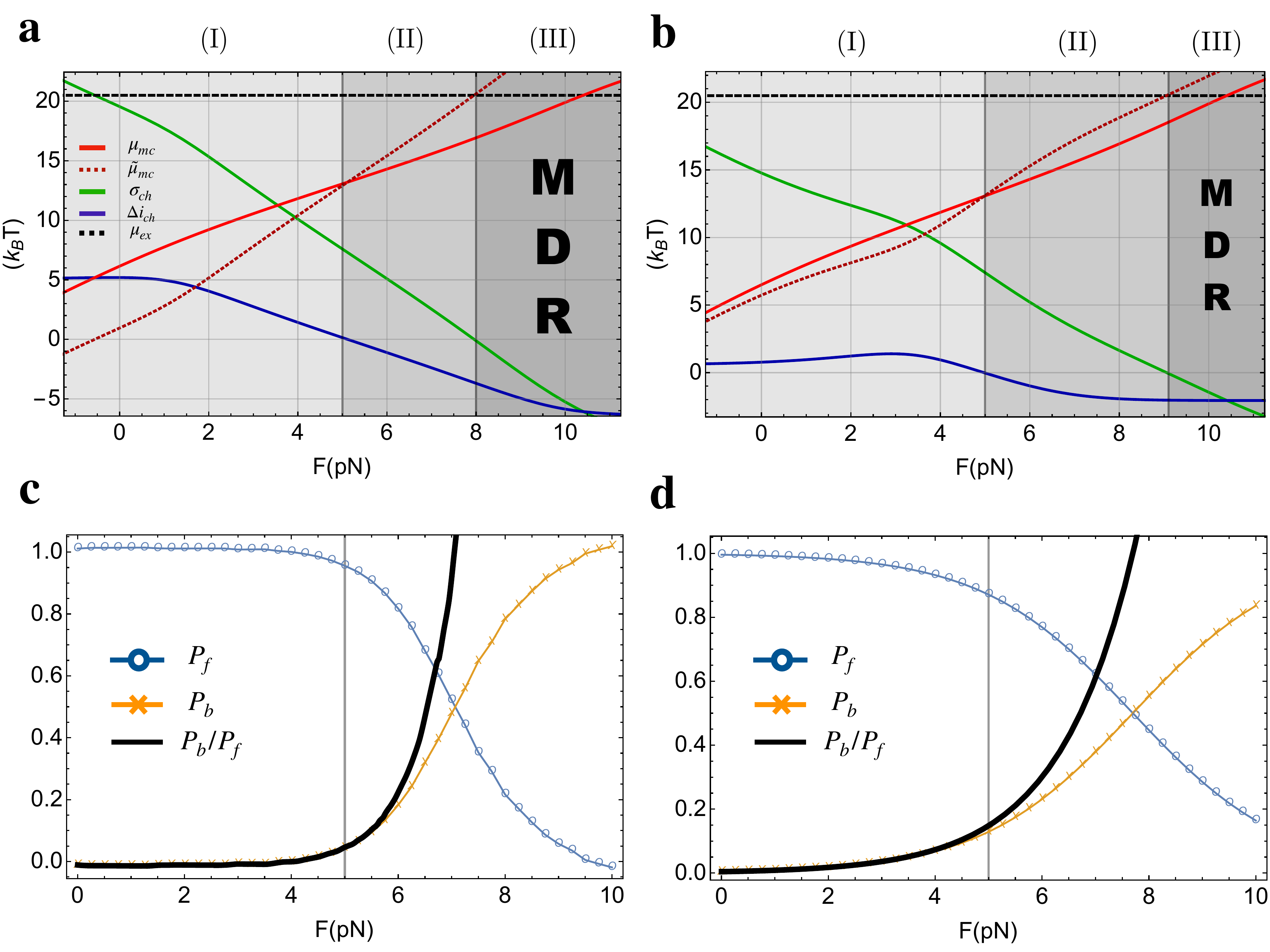}
\caption{\label{Fig:main} Analysis of the experimental data using theory. {\bf (a)-(b)} Contribution of each term in Eq.(\ref{Eq:Infgain}) and Eq.(\ref{Eq:appamu_ch}); $\mu_{mc}$ is the energy needed for driving the mechanical step, $\sigma_{ch}$ is the apparent entropy production in the chemical transitions, $\Delta i_{ch}$ is the change in the PMI during chemical transitions, and $\mu_{ex}$ =  20.5\ $k_BT$ is the input chemical energy~\cite{PhysRevLett.121.218101}. The apparent energy expended to drive the chemical transitions is $\tilde{\mu}_{ch}$. The gray colors and the corresponding numbers are the three regimes explained in the text. MDR stands for Maxwell demon regime. {\bf (a)} [T]=1mM, [D]=0.1mM, and [P$_i$]=1mM. {\bf (b)} [T]=10$\mu$M, [D]=1$\mu$M, and [P$_i$]=1mM. In both {\bf (a)} and {\bf (b)} the regime corresponding to $\mu_{mc}$ exceeding $\mu_{ex}$ is physically not allowed. Other processes (different cycles) would be operative before regime (III) is realized. {\bf (c)} The probability for the forward step ($P_f$; blue), the backward step ($P_b$; yellow), and the fraction ($P_b/P_f$; black), calculated from Ref.~\cite{hwang2018energetic} (see Appendix~\ref{sec:PfPb}). $F=5\pN$ is highlighted by the gray line where $\Delta i_{ch} =0$.
{\bf (d)} Same as {\bf (c)} except the stepping probabilities are calculated using a theory developed elsewhere~\cite{takaki2019kinesin}.
}
\end{figure*}
\subsection{\label{}Cooperative Information Flow ceases as backward step probability increases}

In Fig.\ref{Fig:main}, we plot the $F$-dependent information flow, $\Delta i_{ch}$, and the apparent entropy production $\sigma_{ch}$ during the chemical transitions. The plot naturally divides itself into three regimes depending on the signs of $\Delta i_{ch}$ and $\sigma_{ch}$. (I) $F\lesssim5$pN where $0\leq \Delta i_{ch}$ and $0\leq \sigma_{ch}$. In this force range, there is positive cooperativity between the TH and the microtubule bound LH, thus making the gating process efficient, which in turn results in the motor walking processively on the microtubule . As a consequence, $P_b$, backward step probability  along this pathway is small ($P_b \le 0.05$). (II) In the range $5\text{pN}\lesssim F\lesssim 8\pN-9\pN$, $\Delta i_{ch}\leq 0$ and $0\leq \sigma_{ch}$. At $F=5\text{pN}$, $\Delta i_{ch}$ vanishes (Fig.\ref{Fig:main}a and b). Not coincidentally,  $P_b$ starts to increase rather steeply at $F\approx5\text{pN}$ (Fig.\ref{Fig:main}c and d).  Interestingly, the value of   $F$ at which $\Delta i_{ch}=0$ is {\it independent of the nucleotide concentrations}, as can be noted by comparing Fig.\ref{Fig:main}(a) and Fig.\ref{Fig:main}(b).  (III) In the force range $8\pN-9\pN\lesssim F $ both $\Delta i_{ch} \leq 0 $ and $\sigma_{ch} \leq 0$.  Note that at $F > 10.5$pN the direction of the stationary flow reverses leading to the theoretical possibility of ATP synthesis with $\mu_{mc} > \mu_{ex}$. Although ATP synthesis, with slow rate for kinesin is possible \cite{Hackney05PNAS}, the backward step by ATP synthesis that requires hydrolysis and release of products has not been observed. Hence,  this regime is excluded in our analysis. 

The results in Fig.\ref{Fig:main} along with Eqs.(\ref{Eq:mu_chem})-(\ref{Eq:flux}) allow us to make the following observations. In regime (I), $\Delta i_{ch}$ is positive, which shows that gating is most effective, and power generation is substantial ($\dot{W}$; see Fig.\ref{Fig:CErate}). Interestingly, Hwang and Hyeon have shown using the  thermodynamic uncertainty relation~\cite{hwang2018energetic} that transport efficiency is optimized in this force range.  In regime (II), the information flow is less than optimal. This regime ($\approx$5pN) coincides with the initiation of backward stepping by kinesin, as shown in Fig.\ref{Fig:main}c and d. Thus, the loss of efficient communication between the motor heads ($\Delta i_{ch} <0$) leads to kinesin taking backward steps (the leading head could detach prematurely and goes towards the minus end of microtubule by hydrolyzing ATP).  Fig.~\ref{Fig:main}c and d show the probability of forward step ($P_f$), backward step ($P_b$), and the ratio ($P_b/P_f$) calculated from the previous studies studies~\cite{hwang2018energetic, takaki2019kinesin} (see Appendix~\ref{sec:PfPb}). Despite the different kinetic models used in the two studies (6-states double cycle network for \cite{hwang2016quantifying} and two-states  one dimensional random walk in \cite{takaki2019kinesin}), it is clear that the effect of backward steps become prominent   at $F\approx5\pN $.  In regime (III), $\sigma_{ch}$ is negative, which we refer to as the Maxwell demon regime. The finding that $F_s$ is in regime (II), explains why  kinesin fails to operate under high external load with  positive velocity (Fig.\ref{Fig:vel_fit}).  Finally, at values of $F \leq F_{max}$ inequality $\Delta i \le \sigma$ is satisfied (the green curve lies above the blue curve in Fig.\ref{Fig:main}). Therefore, entropy production [Eq.(\ref{Eq:Epprod})] is always positive while the apparent entropy production takes negative values.


Even though it is difficult to infer the physical or structural meaning of the motor function in the Maxwell Demon regime, it is possible to interpret its origin in terms of thermodynamics. Let us define $\tilde{\mu}_{ch}$ and $\tilde{\mu}_{mc}$ as,
\begin{align} 
\begin{split}
\label{Eq:appamu_ch}
\tilde{\mu}_{ch}= \sigma_{ch},\\ 
\end{split} 
\end{align}
\begin{align} 
\begin{split}
\label{Eq:appamu_mc}
\tilde{\mu}_{mc}=\mu- \sigma_{ch}.\\ 
\end{split} 
\end{align}
The above two equations, Eq.(\ref{Eq:appamu_ch}) and Eq.(\ref{Eq:appamu_mc}), should be compared with Eq.(\ref{Eq:mu_chem}) and Eq.(\ref{Eq:Infgain}), respectively; $\tilde{\mu}_{ch}$ ($\tilde{\mu}_{mc}$) is the apparent energy expended for chemical (mechanical) transition if we have no knowledge of inter-head communication $\Delta i_{ch}$. We can imagine such a scenario  for dimeric motors if one can only access the chemical transitions in one head. If we construct the network of chemical states for the dimeric motor by simply integrating the observation from single motor head, it would lead to the apparent energy transduction given in Eq.(\ref{Eq:appamu_ch}) and Eq.(\ref{Eq:appamu_mc}). In Fig.\ref{Fig:main}, $\tilde{\mu}_{mc}$ reaches and exceeds $\mu_{ex}$ in the Maxwell demon regime. This superficially suggests that the motor could use more energy  to execute the mechanical transition than the input energy $\mu_{ex}$. Thus, without accounting for the inter-head communication, $\Delta i _{ch}$, we would obtain inconsistent energetics  in motors. In this interpretation, the bound head may be thought of the Maxwell demon in the sense that it has information about the trailing or diffusing head. This could be the thermodynamic interpretation of the Maxwell demon.

\section{Myosin V}
\begin{figure*}[]
\centering
\includegraphics[width=\textwidth]{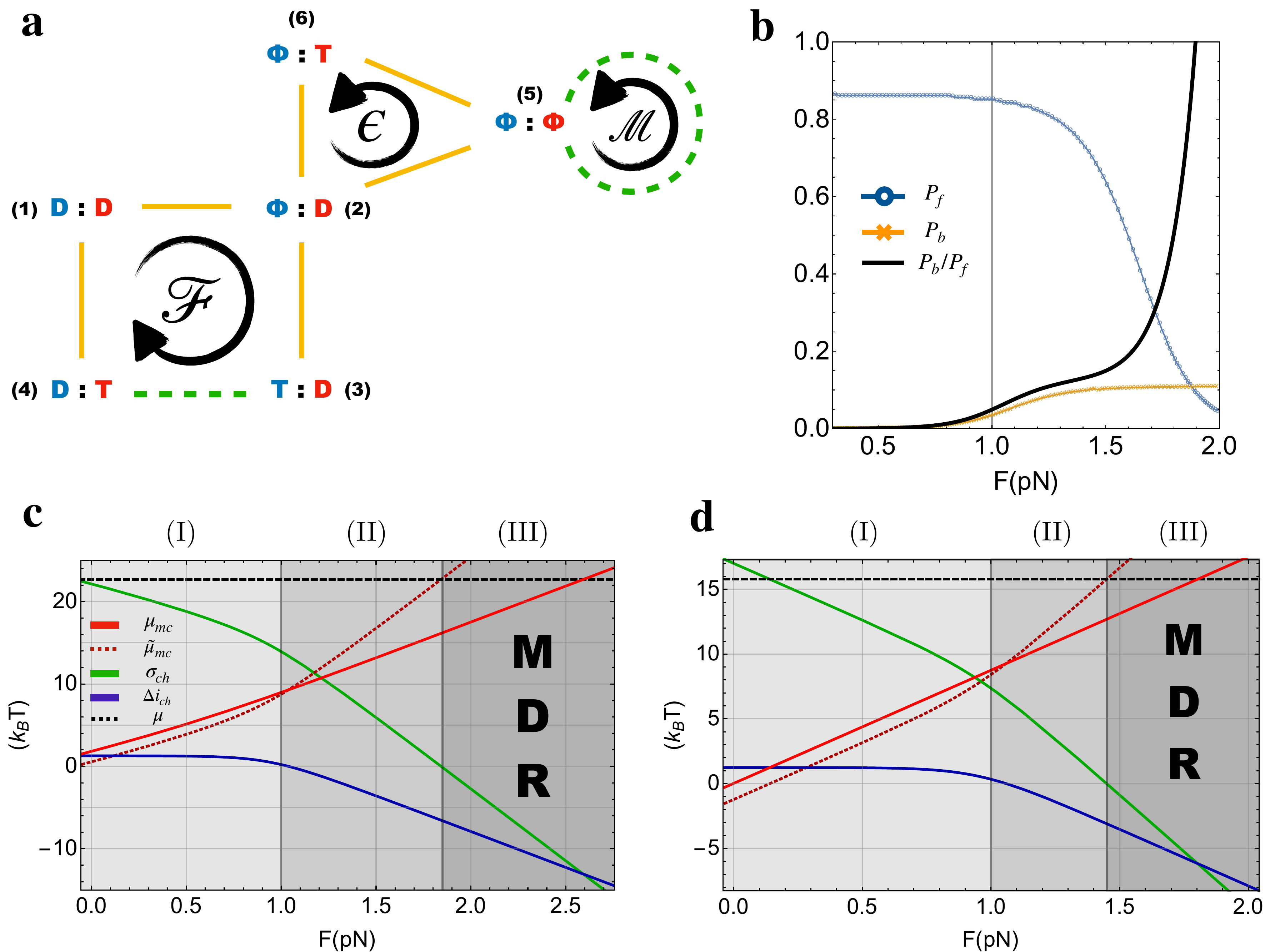}
\caption{\label{Fig:main_MyoV} {\bf (a)} Chemo-mechanical cycle for myosin V. The diagram shows three cycles, $\mathcal{F}$, $\mathcal{\epsilon}$, and $\mathcal{M}$. $\mathcal{F}$ is associated with  stepping utilizing chemical energy by ATP hydrolysis; $\mathcal{\epsilon}$ is a futile cycle in which ATP is consumed but the motor does not step;  $\mathcal{M}$ is describes  mechanical stepping without ATP hydrolysis. In yellow are the chemical transitions and the transitions in dashed green are mechanical transitions. Blue and red colors describe the chemical states in the two heads of the motor. The cycle is adopted from Ref.~\cite{BIERBAUM20111747,hwang2018energetic}. MDR stands for Maxwell demon regime. {\bf (b)} The probability for the forward step ($P_f$), the backward step ($P_b$), and the fraction ($P_b/P_f$), digitized from Ref.~\cite{hinczewski2013design}. The gray line highlights $F\approx 1\pN$ where $\Delta i_{ch}$ becomes negative. Note that at all values of $F$, $P_f + P_b \ne 1$ because of there are pathways associated with stomping of the LH and TH~~\cite{hinczewski2013design}. {\bf (c)-(d)} Plots for the terms in Eq.(\ref{Eq:Infgain}) and  Eq.(\ref{Eq:appamu_ch}) for myosin V. $\mu_{mc}$ is the chemical energy allocated to the mechanical stepping, $\sigma_{ch}$ is the apparent entropy production during chemical transitions, $\Delta i_{ch}$ is the change of PMI during chemical transitions, and $\mu$ is the chemical energy input at given nucleotide concentrations. $\tilde{\mu}_{ch}$ is the apparent energy spent for chemical transitions. Gray regions and the corresponding numbers are the three regimes explained in the text. \{[T]=1$\mMol$ [D]=70$\uMol$ [P]=1$\mMol$\} and \{[T]=1$\uMol$ [D]=70$\uMol$ [P]=1$\mMol$\} for {\bf (c)} and {\bf (d)}, respectively. The regimes in which $\mu_{mc}$ exceeding $\mu_{ex}$ cannot occur (ATP synthesis). Regime (III) would be prevented in practical by enhanced probabilities of other cycles (for instance $\mathcal{\epsilon}$ and $\mathcal{M}$).
}
\end{figure*}

In order to establish that gating, controlled by information flow between the motor heads is applicable to other motors, we next considered  the chemo-mechanical network for Myosin V (Fig.\ref{Fig:main_MyoV}a), a motor that walks also by a hand over hand mechanism on F-actin by taking $36\nm$ steps. The chemo-mechanical model with three cycles was proposed by Bierbaum and Lipowsky~\cite{BIERBAUM20111747}.  The energetic cost, precision, and efficiency associated with the network were investigated recently by Hwang and Hyeon~\cite{hwang2018energetic}.   The cycle, $\mathcal{F}$, is the one that requires chemical coordination leading to the mechanical step. The cycle $\mathcal{\epsilon}$ is the futile cycle that consumes energy but does not lead to a mechanical step. The cycle $\mathcal{M}$ is spontaneous stepping without chemical reactions. 

Our focus is on the information flow in the cycle $\mathcal{F}$ in which mechanical transition requires the coordination of the chemical states between the two motor heads.    
Because experimental data for the concentration of ADP and phosphate for myosin V are not available, we set [D]=70$\uMol$ and [P$_i$]=1$\mMol$, which is appropriate for the {\it in vivo} condition~\cite{hwang2018energetic}. We analyzeed the high ATP concentration [T]=1$\mMol$ and low ATP concentration [T]=1$\uMol$. The nucleotide concentrations \{[T]=1$\mMol$ [D]=70$\uMol$ [P$_i$]=1$\mMol$\} gives chemical energy input $\mu \approx22.7 k_BT$, and \{[T]=1$\uMol$ [D]=70$\uMol$ [P$_i$]=1$\mMol$\} gives chemical energy input $\mu \approx 15.8(k_BT)$. These values were calculated  using the relation $\mu=k_BT\ln K_{eq}\text{[T]/([D][P])} $, where $K_{eq}=4.9 \cdot 10^{11} \uMol$ is the corresponding equilibrium constant. The expected ideal stall load is $F_{max} \approx 2.6 \pN$ and $F_{max} \approx 1.8 \pN$ for \{[T]=1$\mMol$ [D]=70$\uMol$ [P$_i$]=1$\mMol$\} and \{[T]=1$\uMol$ [D]=70$\uMol$ [P$_i$]=1$\mMol$\}, respectively. Note that  the two set nucleotide concentrations yield different amount of input energy. 

Just as in kinesin-1 there are three regimes.  (I) $\Delta i_{ch} \geq 0$ and $\sigma_{ch}\geq 0$ if $F\lesssim 1\pN$, (II) $\Delta i_{ch} \leq 0$ and $\sigma_{ch}\geq 0 $ for $1\pN \lesssim F \lesssim  1.4\pN-1.8 \pN$, and (III) $\Delta i_{ch} \leq 0$ and $\sigma_{ch}\leq 0$ for $1.4 \pN-1.8 \pN \lesssim F$ depending on the nucleotide concentrations. In regime (I) (see Fig.\ref{Fig:main_MyoV}c-d) $\Delta i_{ch}$ is positive, suggesting that communication between the motor heads is efficient  resulting in the forward stepping of the motor with the probability of stepping backwards being negligible.  In regime (II), the motor loses efficient inter-head communication ($\Delta i_{ch} \leq 0$) and the amount of available energy for  mechanical transition diminishes, which is indicated by the negative $\Delta i_{ch}$. This may trigger the increased propensity for the Myosin V to step backwards. A previous theoretical study~\cite{hinczewski2013design} showed backward steps start to be relevant at $F\approx 1\pN$ (Fig.~\ref{Fig:main_MyoV}b), which is the value of force at which $\Delta i_{ch}=0$.  
Regime (III) is the high load regime for Myosin V. The resistive load that signals the transition from regime (II) to (III) is in the range of the experimentally observed stall force ($1.6\pN \lesssim F_s\lesssim 3.0\pN$~\cite{hathcock2020myosin,cappello2007myosin,mehta1999myosin,veigel2002gated,kad2008load,uemura2004mechanochemical,gebhardt2006myosin}). Considering that Myosin V is thought to be more efficient than kinesin-1 ($F_{s}\approx F_{max}$), it is possible that Myosin V could operate close to the Maxwell demon regime (III). 

\section{Discussion and Conclusion}
 The language and interpretation from information theory are extended to illustrate the meaning of gating, an elusive but important concept, in molecular motors. In the framework developed here, gating is viewed as information flow between the two heads of the motor. Communication between the motor heads, which is a qualitative description of gating, has been invoked as the basis for processivity in molecular motors.   Our theory quantifies the gating  phenomenon in terms of parameters that characterize the steps in the catalytic cycle of dimeric molecular motors.  The theory is potentially applicable to a broad class of biological machines. 

We first derived an equality relating the change in PMI to heat dissipation along a transition in a driven kinetic network. By using our theory, we quantified gating in kinesin and Myosin V.   The two most important findings are the following. First, we showed that gating is cooperative, at resistive forces less than a critical value, $F_c$, which means that communication between the two heads is effective. As a consequence, the chemical transitions results in the trailing head detaching with much greater probability than the leading head. In this case, the probability that the motor takes a forward step far exceeds the probability of backward steps.

Second, as the magnitude of the force increases, there is a loss in gating efficiency, and  information $\Delta i_{ch}$ becomes negative. The transition between low force positive cooperative interaction between the motor heads and loss of inter-head communication, due to negative cooperativity,  occurs at $F_c$ where $\Delta i_{ch}=0$. At $F_c$ there is a lack of correlation (or communication)  between the two heads. Surprisingly, for both the motors the values of $F_c$, which is independent of the input energy, coincides  when the probability for backward transition starts to increase. 

Our findings show that despite the significant differences in the chemical cycle and the architecture of kinesin-1 and Myosin V, we obtain similar results with regard to the force-dependent information flow, and hence gating. This is not a coincidence and that this class of  motors have evolved to work efficiently at forces less than $F_c$. Efficient gating could be the universal working principle in dimeric molecular motors, which means that $F$ has to be $F_c$ under {\it in vivo} conditions.

We are now in a position to  provide an interpretation of positive information flow, which results in kinesin taking multiple steps towards the plus end of the MT without detaching. The coordination between motors (see Fig.\ref{Fig:schematics}c) may be quantified using,
\begin{equation}
\Delta i = k_B \Big(\ln \frac{p(x_5,y_5)}{p(x_5)p(y_5)} - \ln \frac{p(x_1,y_1)}{p(x_1)p(y_1)}\Big).
\label{info}
\end{equation}
Positive $\Delta i$ means that the forward mechanical step, [(5)$\rightarrow$(1) in Fig.\ref{Fig:schematics}c], occurs with high probability. For this value to be large and positive, which implies effective communication, the chemical states of the two heads must be more correlated in state $5$ (before the mechanical step) than in state $1$ (after the step). 
In other words, $\frac{p(x_5,y_5)}{p(x_5)p(y_5)} > \frac{p(x_1,y_1)}{p(x_1)p(y_1)}$.
The reason for this is intuitive: once the chemical state before the mechanical step is reached, it does not behoove the motor to leave that state, and thus it is beneficial to have strong correlation between the chemical states of the two heads.
In contrast, after the mechanical step is completed (state $1$), the chemical state of the two heads should be loosely correlated, in order to immediately leave the landing state, which is amenable to revert the step.
We note that this is realized in molecular motors: upon completing a mechanical step by binding to the track, kinesin and myosin release ADP and phosphate, respectively, which consolidates the two-head-bound state, which  in the framework of this paper means that $\frac{p(x_1,y_1)}{p(x_1)p(y_1)}$ should be small.

Our analysis explains the reason for the apparent inefficiency of kinesin~\cite{PhysRevLett.121.218101}.  For mechanical stepping to happen [(5)$\rightarrow$(1) in Fig.\ref{Fig:schematics}c], kinesin has to first undergo chemical transitions [(1)$\rightarrow$(5) in Fig.\ref{Fig:schematics}c]. If the motor were to take a forward step, certain amount of energy has to be expended to complete the chemical transitions at all values of $F$. At forces exceeding $\approx 5\pN$, there is a breakdown in inter-head communication. In addition, the work required to complete the step increases dramatically (see Fig.\ref{Fig:main}a). As a result of loss of communication between the two heads and the increase in the mechanical energy to complete a step, other pathways (in particular backward step) become prominent. Thus, above $F_c$ the motor efficiency is inevitably compromised. 
Although an experiment similar to the one for kinesin~\cite{PhysRevLett.121.218101} have not been performed for Myosin V, our prediction is that the results would be similar (Myosin V would be most efficient if $F \le F_c \approx 1\pN$).  

In contrast to dimeric motors in which substantial potion of the input energy is dissipated, experiments show that for the ATP-synthesizing rotary motor $\text{F}_1$-ATPase operates at nearly 100\% efficiency~\cite{Toyabe_2010}. A plausible explanation for this high efficiency is that the chemical transitions of $\text{F}_1$-ATPase is fully coupled to the mechanical transitions.  The reversible chemical-mechanical coupling is markedly different from processive dimeric motors.

From a structural perspective, information flow between the heads must be linked to action at a distance expressed in terms of allostery~\cite{thirumalai2019symmetry}. In kinesin, that walks on microtubule, the two heads are about 8nm apart whereas the distance between the motor heads in the F-actin bound Myosin V is about 36nm. How, $\Delta i_{ch}$,  the information flow is linked to the structural changes in the motor driven by binding and  hydrolysis of ATP followed by ADP release is unclear. The molecular link needs to be established before the design principles in naturally occurring motors can be fully understood.  

\bigskip

\noindent{\bf Acknowledgements:} We are grateful to Jordon Horowitz for several useful comments on an earlier version of the manuscript. This work was supported by NSF (CHE 19-00093), NIH (GM - 107703) and the Welch Foundation Grant F-0019 through the Collie-Welch chair.

\appendix
\section{\label{sec:expressions}Expressions for the key quantities}
{\bf Kinesin-1:} 
In this Appendix, we enumerate various quantities in our theory that are needed  to analyze the kinetic network for kinesin-1 shown in Fig.\ref{Fig:schematics}c. To stress the steady state probability, we denote the probability being in state $i$ as $p^s_i$ ($i$ spans from 1 to 5 for the network in Fig.\ref{Fig:schematics}c). Heat dissipation is expressed using the transition rate from state $i$ to state $j$, $w_{ij}$. We identify the trailing head (blue in Fig.\ref{Fig:schematics}c) as $\Y$ and the leading head (red in Fig.\ref{Fig:schematics}c) as $\X$. For the kinesin-1 network in Fig.~\ref{Fig:schematics}c we obtain, 
\begin{align} 
\label{eq:kin1_key}
&\Delta i_{\X,ch} = k_B\ln\frac{p_2^s}{p_4^s} & &\Delta i_{\Y,ch} = k_B\ln\frac{p_4^s(p_1^s+p_2^s)}{(p_4^s+p_5^s)p_2^s}\\
&\Delta s_{\X,ch} = k_B\ln\frac{p_1^s}{p_5^s} & &\Delta s_{\Y,ch} = k_B\ln\frac{p_1^s+p_2^s}{p_4^s+p_5^s} \\
&\Delta s_{\XY,ch}^\X = k_B\ln\frac{p_1^s p_4^s}{p_2^s p_5^s} &  &\Delta s_{\XY,ch}^\Y = k_B\ln\frac{p_2^s}{p_4^s}\\
&\Delta i_{ch} = k_B\ln\frac{p_1^s+p_2^s}{p_4^s+p_5^s} &  &\Delta s_{\XY,ch} = k_B\ln\frac{p_1^s}{p_5^s}.
\end{align}

\begin{align} 
\label{}
&Q_{\X,ch} =k_B \ln\frac{w_{12}w_{45}}{w_{21}w_{54}} & & Q_{\Y,ch} = k_B\ln\frac{w_{23}w_{34}}{w_{32}w_{43}}\\
&Q_{ch} = k_B\ln\frac{w_{12}w_{23}w_{34}w_{45}}{w_{21}w_{32}w_{43}w_{54}} & &  Q_{mc} = k_B\ln\frac{w_{51}}{w_{15}}.
\end{align}

{\bf Myosin V:}
The trailing head for Myosin V is in blue, and is denoted as $\Y$, and the leading head in red corresponds to $\X$ (see Fig.\ref{Fig:main_MyoV}a). For the chemical transitions associated with $(4)\rightarrow(1)\rightarrow(2)\rightarrow(3)$ in Fig.\ref{Fig:main_MyoV}a, we obtain the following expressions.
\begin{align} 
\label{}
&\Delta i_{\X,ch} = k_B\ln\frac{p_1^s(p_4^s+p_6^s)}{(p_1^s+p_2^s+p_3^s)p_4^s} & &\Delta i_{\Y,ch} = k_B\ln\frac{p_1^s+p_4^s}{p_1^s}\\
&\Delta s_{\X,ch} =k_B \ln\frac{p_4^s+p_6^s}{p_1^s+p_2^s+p_3^s} & & \Delta s_{\Y,ch} = k_B\ln\frac{p_1^s+p_4^s}{p_3^s}\\
&\Delta s_{\XY,ch}^\X = k_B\ln\frac{p_4^s}{p_1^s } & & \Delta s_{\XY,ch}^\Y = k_B\ln\frac{p_1^s}{p_3^s}\\
&\Delta i_{ch} = k_B\ln\frac{(p_1^s+p_4^s)(p_4^s+p_6^s)}{(p_1^s+p_2^s+p_3^s)p_4^s} & & \Delta s_{\XY,ch} = k_B\ln\frac{p_4^s}{p_3^s}.
\end{align}

\begin{align} 
\label{}
&Q_{\X,ch} = k_B\ln\frac{w_{41}}{w_{14}} & &  Q_{\Y,ch} = k_B\ln\frac{w_{12}w_{23}}{w_{21}w_{32}}\\
&Q_{ch} = k_B\ln\frac{w_{12}w_{23}w_{41}}{w_{21}w_{32}w_{14}} & &  Q_{mc} = k_B\ln\frac{w_{34}}{w_{43}}.
\end{align}

Expressions for $\sigma$ and $\rho$ can be obtained using the following relations. 
\begin{align} 
\begin{split}
&\sigma_{\X,ch} =  Q_{\X,ch} + \Delta s_{\X,ch}\\
&\sigma_{\Y,ch} =  Q_{\Y,ch} + \Delta s_{\Y,ch}\\
&\sigma_{ch} = Q_{ch}+\Delta s_{\X,ch} + \Delta s_{\Y,ch}.\\
\\
&\rho_{\X,ch} = \sigma_{\X,ch} - \Delta i_{\X,ch}\\
&\rho_{\Y,ch} = \sigma_{\Y,ch} - \Delta i_{\Y,ch}\\
&\rho_{ch} = \sigma_{ch} - \Delta i_{ch}.\\
\end{split} 
\end{align}

\section{\label{sec:parameters}Kinetic rates for kinesin-1}
In order to extract the parameters characterizing the catalytic cycle (Fig.\ref{Fig:schematics}c), we simultaneously fit the two average velocities as a function of force at the nucleotide concentrations, \{[T]=1mM, [D]=0.1mM, [P]=1mM\} and \{[T]=10$\mu$M, [D]=1$\mu$M, [P]=1mM\}~\cite{PhysRevLett.121.218101}. We ignored the negative points, since we only model the forward stepping cycle. Following the previous study~\cite{liepelt2007kinesin}, we use the functional form (Bell model) for the mechanical transition rates, $w_{51}=w^o_{51}\text{e}^{- \theta f d_0/k_B}$ and $w_{15}=w^o_{15}\text{e}^{(1-\theta) f d_o/k_B}$. The load acting on the motor, $F$, modifies the bare rates for the forward step and backward steps, $w^o_{51}$ and $w^o_{15}$, respectively, depending on the load distribution factor $\theta$. Note that the range of $\theta$ is $0 \leq \theta \leq 1$. 

Liepelt and Lipowsky~\cite{liepelt2007kinesin} modified the  chemical transition rates in order to account the force dependence by using,  
\begin{equation}  
w_{ij}=[\text{S}]2w_{ij}^o(1+\text{e}^{\chi_{ij} f d_o/k_B})^{-1},
\label{chemtrans}
\end{equation}
where $\chi_{ij}=\chi_{ji}\geq0$ is a dimension-less force parameter and $d_o=8$ nm is the step size of kinesin. In Eq. (\ref{chemtrans}), [S] represents the nucleotide concentrations ([T],[D],[P]) if a specific  transition is driven by  binding of ATP, ADP, or P$_i$. For example, the (4)$\rightarrow$(5) is driven by ATP binding, and hence [S] would correspond to the concentration [T] of ATP. We emphasize that the reverse transitions $(2) \rightarrow (1)$ and $(4) \rightarrow (3)$ in Fig.\ref{Fig:schematics}c requires the binding of ADP and phosphate, respectively. Note that $\ln \frac{w_{ij}}{w_{ji}}$ in chemical transitions are not $F$-dependent because $\chi_{ij}=\chi_{ji}$, which suggests that dissipation from the chemical transitions is not affected by $F$. These functional forms of the rates enable us to separate the catalytic cycle of the motor into chemical transitions and mechanical transition~\cite{liepelt2007kinesin,hwang2018energetic}.

\begin{table}[]
\begin{center}
\resizebox{\columnwidth}{!}{%
  \begin{tabular}{|c | c|| c | c|| c |c|}
    \hline
    \text{Parameter} &\text{Value} & \text{Parameter} &\text{Value} & \text{Parameter} &\text{Value}\\
    \hline \hline
    $w^o_{12}$      &$384.4(s^{-1})$ &$w^o_{21}$      &$1.5~(s^{-1}\mu \text{M}^{-1})$ &$\chi_{12}$   &$0.4$\\ 
    $w^o_{23}$    &$371.2(s^{-1})$  &$w^o_{32}$    &$33.2(s^{-1})$ & $\chi_{23}$	&   $0$\\
    $w^o_{34}$ &$459.5(s^{-1})$  &$w^o_{43}$ &$2.7(s^{-1}\mu \text{M}^{-1})$ &$\chi_{34}$	&   $0$\\
    $w^o_{45}$	 &$29.8(s^{-1}\mu \text{M}^{-1})$  &$w^o_{54}$	 &$94.7(s^{-1})$ &$\chi_{45}$	&   $0$\\
    $\ ^\dagger w^o_{51}$ &$3\times10^5(s^{-1})$ &$w^o_{15}$	 &$0.7(s^{-1})$  &$\theta$&   $1$\\
    \hline
  \end{tabular}
}
\end{center} 
\caption{\label{Table:parameters}Extracted parameters by simultaneous fitting to the experiment~\cite{PhysRevLett.121.218101}. $\dagger$ : We use the value $w^o_{51}$ given elsewhere~\cite{liepelt2007kinesin}.}
\end{table}

\begin{figure}[htbp]
\begin{center}
\includegraphics[width=0.5\textwidth]{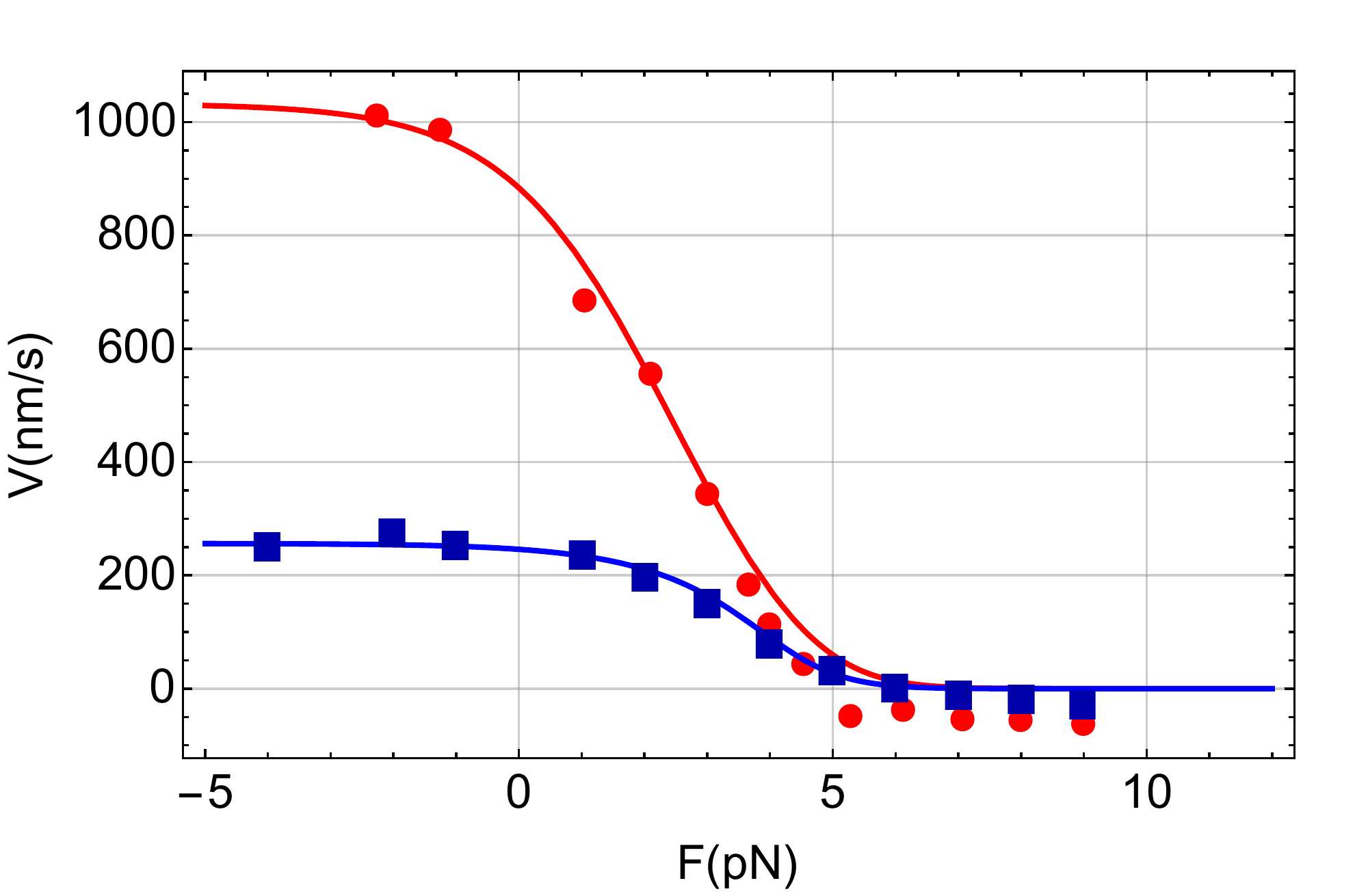}
\end{center}
\caption{\label{Fig:vel_fit} Simultaneous fit to the average kinesin-velocity (= $J_{mc} \Delta X$ where $\Delta X$ is the motor step size) as a function of external load at different concentrations of ATP, ADP and phosphate. Red dots and blue squares are from the experiment~\cite{PhysRevLett.121.218101} for \{[T]=1mM, [D]=0.1mM, [P]=1mM\} and \{[T]=10$\mu$M, [D]=1$\mu$M, [P]=1mM\}, respectively. The red and blue solid curves are the fits using the kinetic network in Fig.\ref{Fig:schematics}c. }
\end{figure}

\section{\label{sec:PfPb} Calculations for $P_f$ and $P_b$}
A key finding in our work is that gating or information flow between the motor heads ceases to be efficient or does not occur at $F$ values at which the probability, $P_b$, starts to increase. At this force value $\Delta i_{ch}$ vanishes for both Kinesin-1 and Myosin V. To demonstrate that this is the case, we calculated the probabilities for forward  ($P_f$) and backward step ($P_b$) for kinesin-1 and Myosin V using the results from the previous studies~\cite{hwang2018energetic,takaki2019kinesin} and \cite{hinczewski2013design}, respectively. Plots in \ref{Fig:main_MyoV}b were obtained by digitizing the results in Fig.1D in \cite{hinczewski2013design}, where analytic expressions for $P_f$ and $P_b$ as a function of $F$ are derived. For Kinesin-1 plotted in Fig.\ref{Fig:main}c, we first digitized Fig.1B in \cite{hwang2018energetic}, then calculated $P_f=J_F/(J_F+J_B)$ and $P_b=J_B/(J_F+J_B)$, where $J_F$ and $J_B$ are the fluxes along the forward step and backward step, respectively. Similarly, the results in Fig.\ref{Fig:main}d, were computed using  $P_f=J^+ /(J^+ +J^-)$ and $P_f=J^- /(J^+ +J^-)$.  The fluxes for forward s ($J+$) and backward steps ($J^-$) were calculated using Eq.[2] in \cite{takaki2019kinesin}.     

\bibliography{ref.bib,Kinesin.bib}
\end{document}